\documentclass[twocolumn, 12pt]{article}
\usepackage[utf8]{inputenc}
\usepackage[english]{babel}
\usepackage{graphicx}
\usepackage{geometry}
\usepackage{multirow}
\usepackage{array}
\usepackage{prettyref}
\usepackage{booktabs}
\usepackage{units}
\usepackage{amsmath}
\usepackage{amssymb}
\usepackage{wasysym}
\usepackage{subscript}
\usepackage{textcomp}
\usepackage{gensymb}
\usepackage{xcolor}
\usepackage{hyperref}

\usepackage{float}
\usepackage[export]{adjustbox}

\usepackage{authblk}
\usepackage[font=footnotesize,labelfont=bf]{caption}
\title{Laser-cooled ytterbium ion microwave frequency standard}

\author[1,2]{S. Mulholland\footnote{current email sean.mulholland@jpl.nasa.gov}}
\author[1]{H. A. Klein}
\author[1]{G.P. Barwood}
\author[1]{S. Donnellan}
\author[1]{D. Gentle}
\author[1]{G.\,Huang}
\author[1]{G. Walsh}
\author[2]{P.E.G. Baird}
\author[1,2]{P. Gill}

\affil[1]{National Physical Laboratory, Teddington, TW11 0LW, UK}
\affil[2]{Clarendon Laboratory, University of Oxford, Oxford, OX1 3PU, UK}
\date{}                     %% if you don't need date to appear
\begin{document}
\sloppy

\maketitle
\begin{abstract}
We report on the
development of a trapped-ion, microwave frequency standard based on the 12.6\,GHz hyperfine transition in
laser-cooled ytterbium-171 ions. The entire system fits into a 6U
19-inch rack unit ($\unit[51\times49\times28]{cm} $) and comprises laser, electronics and
physics package subsystems.  As a first step towards a full evaluation of the system capability, we have measured the frequency instability of our system which is  $3.6\times10^{-12}/\surd\tau$
for averaging times between $\unit[30]{s}$ and $\unit[1500]{s}$.

%\fussy

\end{abstract}

\section{\label{sec:level1}Introduction}

The current state of the art of portable microwave clocks and frequency standards are based on beam-tubes \cite{Cutler2005}, vapor cells \cite{camparo207,MicrosemiInc.2017}, 
and trapped ions in buffer gas \cite{Jau2012,Schwindt2016,Tjoelker2016}.
In beam-tube clocks the linewidth of the atomic resonance is proportional to the length of the beam-tube and this places limits to the extent to which they can be miniaturized. In addition, the atom-flux in a beam-tube is thermal which limits performance by the second-order Doppler effect, and the magnets that provide state-selection produce magnetic inhomogeneities along the beam path that result in a difficult to characterize second-order Zeeman shift.
Vapor-cell clocks are sensitive to the cell temperature and pressure, their output frequency drifts on timescales longer than approximately one day, and long-term
aging phenomena affect the frequency \cite{Camparo2005}. Buffer-gas cooled trapped ion systems are susceptible to
collisional and pressure related shifts that limit long-term stability and accuracy \cite{Jau2015,Warrington2002,Park2007a}. The use of thermal atoms in these systems also limits frequency reproducibility. To reduce these effects, the highest-performance laboratory clocks, such as cesium fountains \cite{Heavner2014,Guena2012,Szymaniec2010}  use
laser cooling in an ultra-high vacuum (UHV).

\begin{figure}
\includegraphics[ width=1\columnwidth]{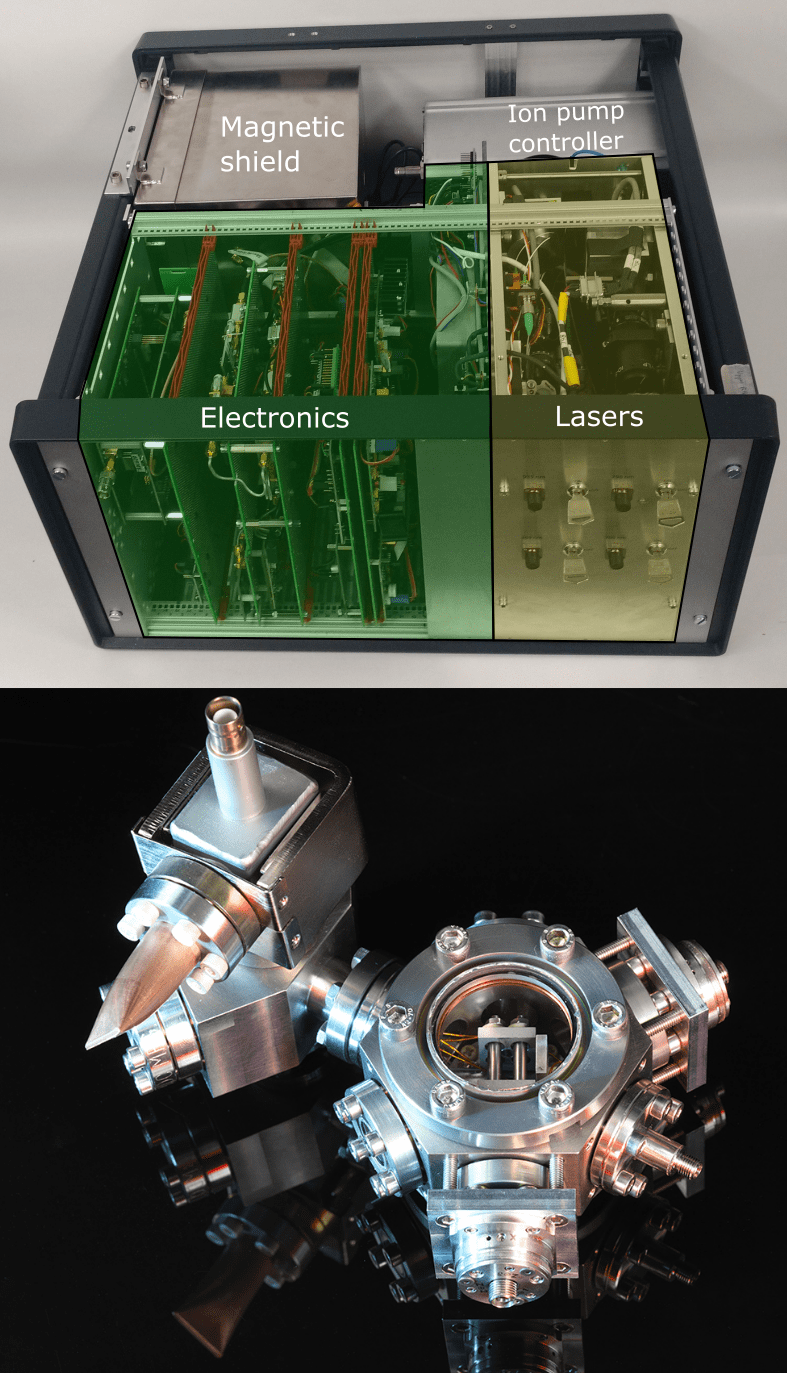}
\caption{ Labelled photograph of the frequency standard within the 6U, 19-inch rack case (top) and the vacuum chamber (bottom). The majority
of the electronics is located on the vertical 6U cards at the front
left of the case. The laser system is housed at the front-right of
the case. The physics package, within a magnetic shield, is located
at the back-left. The ion pump controller is visible at the back-right. A power supply (not shown) is at the rear.
The vacuum chamber houses the ion trap, pumps, and electrical feedthroughs. Not shown on the image are the PMT, imaging optics, and coils that are rigidly mounted to the chamber. Additionally, we fitted a soft iron enclosure around the ion pump to confine the magnetic field from the pole pieces.    
\label{fig:6U}}
\end{figure}

We report on the development of a laboratory prototype system for a  compact, trapped-ion, microwave
frequency standard incorporating laser cooling.  The system is shown in Figure \ref{fig:6U} and is housed in a 6U 19-inch
rack ($\unit[51\times49\times28]{cm} $).

\sloppy
Non-portable laboratory versions of laser-cooled, ytterbium-ion, microwave frequency
standards  have been built \cite{Warrington1999,Phoonthong2014} and
it is estimated that a potential fractional frequency instability of $5\times10^{-14}/\sqrt[]{\tau}$
and fractional frequency uncertainty of $\sim5\times10^{-15}$ are achievable in laboratory systems \cite{Park2007a,Warrington1999}.
Clock applications at this level of performance  include network synchronization, navigation and accurate timing capabilities independent of Global Navigation Satellite Systems (GNSS). 

In this paper, our trapped ion microwave frequency standard is described. The apparatus (physics package, lasers, and electronics) is outlined in section \ref{overview}, the methods of operating the frequency standard are described in section \ref{sec:Meth}. The results achieved to date, and an analysis of the limitations to stability and accuracy of our frequency standard,  are presented in section \ref{results}.

\section{System Overview}\label{overview}

The frequency standard \cite{MulhollandThesis} consists of three main subsystems: Ions are created,
trapped and probed within an evacuated linear ion trap system; all the required lasers
and optics with the exception of a wavemeter are housed in a 6U pullout rack unit. The 6U rack also contains the electronics and microwave system that
runs the operational sequence and locks an oscillator to the atomic
resonance. These systems are detailed in the following subsections.  During normal clock operation, when the oven is not being fired the system consumes a maximum power of only 80\,W.  

\begin{figure}
\includegraphics[ width=1\columnwidth]{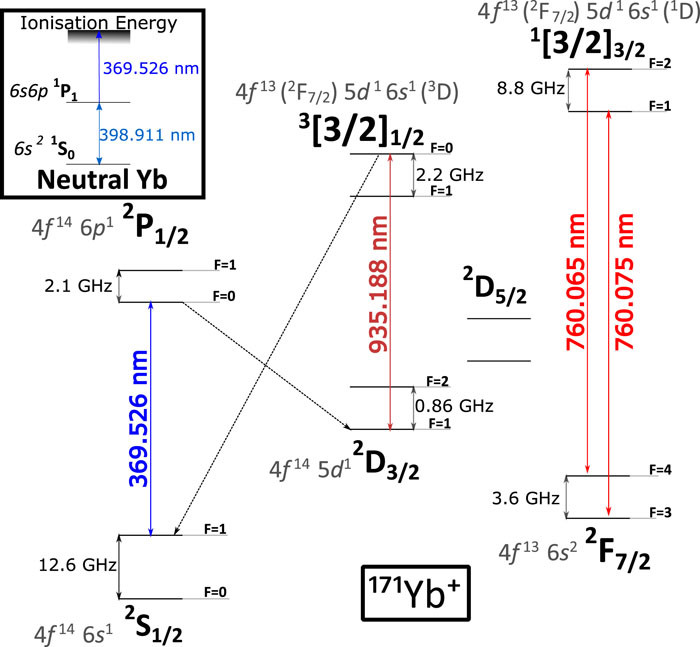} %The previous version of this term scheme contained an error in the upper state of the 760nm transition.
\caption{Partial term schemes showing the driven atomic transitions and the required laser wavelengths and microwave frequencies. The inset shows the transitions in neutral ytterbium used for photoionization. The main diagram shows the $^{171}\mathrm{Yb}^{+}$ term scheme.  \label{fig:TermScheme}}

\end{figure}

The term schemes showing the atomic transitions used in the frequency standard for both
neutral and singly-ionized \textsuperscript{171}Yb are shown in Figure \ref{fig:TermScheme}.

To laser cool the ions, microwaves at 12.6\,GHz are applied and a 369\,nm
laser is tuned to the $\text{F}=1\rightarrow\text{F}'=0$ component
of the $6s: \, ^{2}\text{S}_{\nicefrac{1}{2}}\,\rightarrow\,6p:\,^{2}\text{P}_{\nicefrac{1}{2}}$
transition. To mix the ground states during cooling, the microwave transition is driven rapidly with a Rabi frequency of 100\,kHz. The cooling transition is a strong, dipole-allowed transition
with a natural linewidth, $\gamma_{369}$, of 19.6\,MHz full width at half maximum (FWHM) \cite{Hayes2009}.
The cooling cycle is not completely `closed'   since
the upper state can decay with a probability of about
$\nicefrac{1}{200}$ into a $^{2}\text{D}_{\nicefrac{3}{2}}$ state with a 50\,ms lifetime
\cite{Yu2000a}. To prevent \textquoteleft shelving\textquoteright{}
the ions in this state, a laser of 935\,nm is tuned to the  ${^{2}\mathrm{D}_{\nicefrac{3}{2}}}(\mathrm{F}=1)\,\rightarrow\,{}^{3}[\nicefrac{3}{2}]{}_{\nicefrac{1}{2}}(\text{F}'=0)$ transition.  During laser-cooling the detected ion fluorescence rate of the 369-nm transition is  $\unit[1.5\times10^5]{counts/s}$ from 300 ions, which corresponds to each of the ions scattering approximately $\unit[2\times10^{4}]{photons/s}$  taking into account our collection efficiency of  2.7\%.  

Ytterbium has a low-lying F-state with a natural lifetime of the order
of 10 years  \cite{Roberts1997}. The long lifetime can
be exploited to provide a narrow linewidth for an optical clock  \cite{Huntemann2016,Godun2014}.
However, population trapping in the F-state can occur by a collisional mechanism, and this can degrade the performance of a clock \cite{Jau2015,Fisk1995}. To avoid this we took the precaution of including a 760\,nm laser in our system. Additionally, the frequency standard operates under UHV (typically $<10^{-10}\,\mathrm{mbar}$ is measured by the ion pump controller) and the collisional route into the F-state is suppressed. We did not observe significant population trapping in the F-state, even when operating without the 760\,nm laser.

The laser cooling of \textsuperscript{171}Yb\textsuperscript{+}
has a further complication. In the absence of an external magnetic
field, the ions will become optically pumped in the $^{2}\text{S}_{\nicefrac{1}{2}}\text{, F=1}$
dark states. A magnetic field of 700\,$\mathrm{\mu}\text{T}$ was  applied by a coil  to induce a precession of the ions' dipole moments to destabilize the dark
states \cite{Berkeland2002a}. This large magnetic field must be
turned off during the probe phase of the clock cycle to reduce
the uncertainty in the second-order Zeeman shift. When switched off, the magnetic field falls to $1\%$ of it's full value within $\unit[12]{ms}$, limited by eddy currents.  

The Doppler limit for laser cooling is around $0.5\,\mathrm{mK}$ \cite{Hayes2009,Keller2015}.
However, confining large clouds of ions in a quadrupole linear RF trap inevitably leads to significant micromotion and therefore the ion temperatures were higher than this Doppler limit.

\subsection{\label{sec:level2}Physics package}

The physics package comprises a linear ion trap within a vacuum housing and is mounted inside a mu-metal magnetic shield. A photomultiplier (PMT) with imaging optics, magnetic bias coils, and fiber optic collimators  are attached outside the vacuum chamber.

 \sloppy

\begin{figure}[t]

\begin{centering}
\includegraphics[width=0.75\columnwidth]{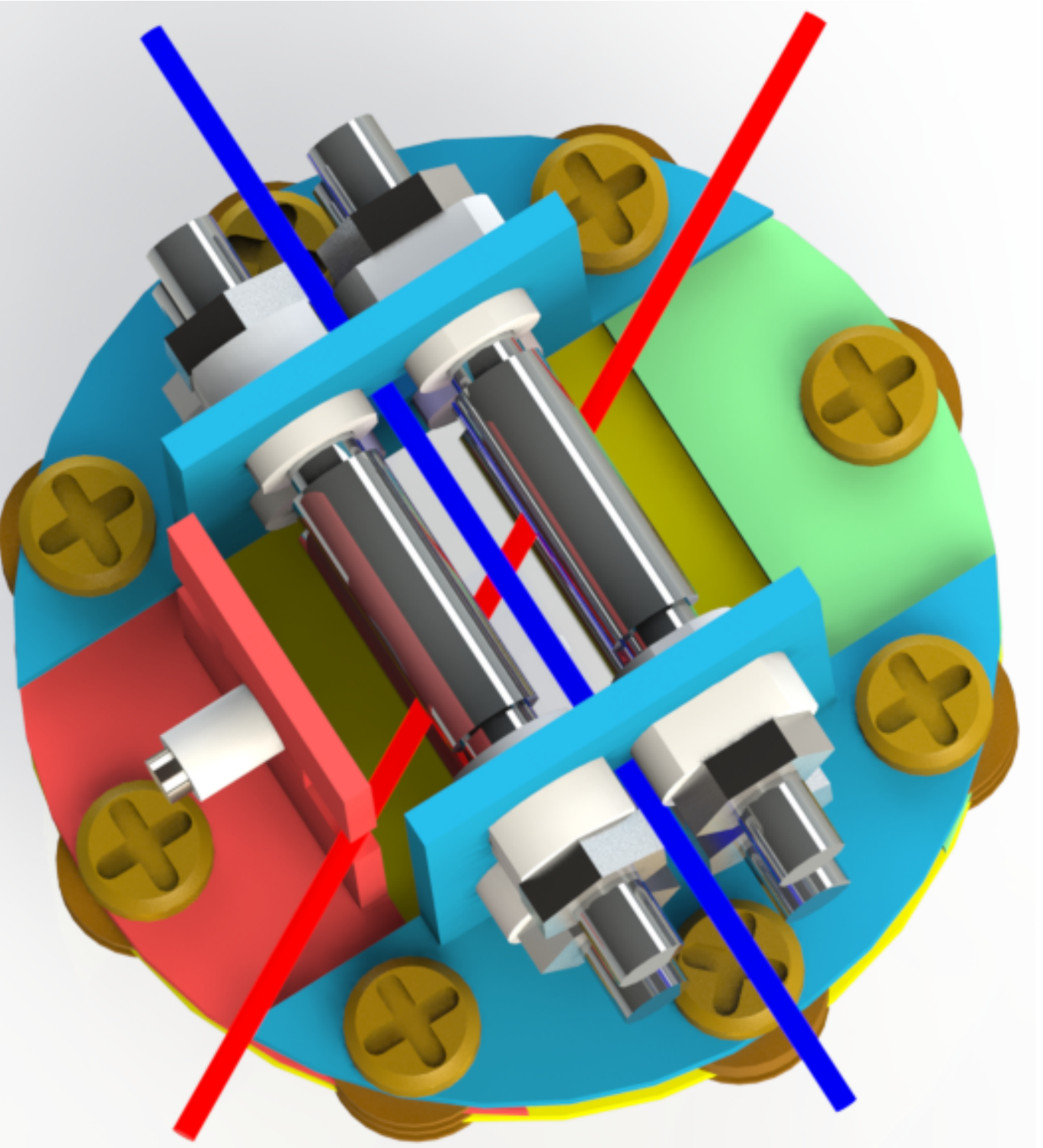}\caption{Three-dimensional illustration of the ion trap with the laser paths shown. The electrodes are
colored for identification: endplates: blue, horizontal compensation: red, vertical compensation: yellow, ground: green. The oven is embedded,  within an alumina insulator,  in the horizontal compensation electrode. The four rods are isolated from the endplates by alumina bushes. The axial laser beam contains the UV wavelengths and passes through holes bored in the endplates. The IR laser beam is 60\degree\,to the axis and passes through a cut-out on the electrode that houses the oven. PEEK washers are used to isolate the plates and PEEK screws attach the ion trap to the vacuum chamber.\label{fig:ion-trap}}
\par\end{centering}
\end{figure}

The ion trap comprises four rods and five plate electrodes as shown in Figure \ref{fig:ion-trap}. The rods provide trapping in the radial direction, and a pair of plates confine the ions axially. Two orthogonal plate electrodes are used to compensate for stray DC fields and minimize radial micromotion. The final plate electrode is grounded. The plates are isolated from each other by polyether ether ketone (PEEK) washers. The trap assembly was screwed into internal mounting holes in the vacuum chamber with PEEK screws. The oven, which provides the source of neutral ytterbium atoms, is embedded within a ceramic housing in an electrode. The front of the oven was spot-welded on the horizontal compensation electrode. The dimensions of the ion trap are summarized in Table \ref{tab:mk2-ion-trap}. The ratio of the rod electrode radius to the ion-electrode separation, ${r_{e}/r_{0}}$ was 0.63, which is less than the value of approximately 1.145 that gives the best approximation to a  quadrupolar  electric field \cite{Denison71,Pedregosa2010}.  Departing from the ideal quadrupolar field  causes an increase in orbits within the ion trap that are unstable, and gain energy from the trapping field \cite{Major2005}.  

 As this device was designed to be transportable, it was essential that the various optical elements be robustly mounted. The laser launchers, PMT and magnetic shields are all directly mounted onto the vacuum chamber. The vacuum chamber was made from titanium as it is non-magnetic and exhibits low outgassing. The fused silica windows were a low-profile design and indium-sealed to titanium flanges. The vacuum system was baked at 130\,\degree C  for a week while connected to a turbo-molecular pump to remove outgassing material sorbed to the chamber walls. Indium seals on the windows limited the baking temperature. After  bake-out, the internal getter pump was activated, and the external pump was valved off and removed. Getter pumps do not pump noble gases so, for long-term operation, an ion pump was also used.

\begin{table}
\centering{}
\caption[Ion trap parameters ]{Ion trap parameters \label{tab:mk2-ion-trap}}
\begin{small}
\begin{tabular*}{1\columnwidth}{@{\extracolsep{\fill}}>{\centering}m{0.3\columnwidth}>{\raggedright}m{0.6\columnwidth}}
\toprule 
\textbf{rod electrodes} & titanium electrodes, radius, $r_{e}=\unit[2]{mm}$.\tabularnewline
\midrule 
\textbf{endcap electrodes} & aluminum plates separated by $\unit[17.8]{mm}.$\tabularnewline
\midrule 
\textbf{rod spacing} & $\unit[7.3]{mm}$ center-to-center.\tabularnewline
\midrule 
\textbf{laser access} & on-axis 4.8\,mm diameter bore through endplates. $3.5\times4$\,mm
notch in oven holder $60\degree$ to axis.\tabularnewline
\midrule 
\textbf{ion-electrode separation, $\boldsymbol{r_{0}}$} & 3.16\,mm.\tabularnewline
\midrule 
\textbf{$\boldsymbol{r_{e}/r_{0}}$} & 0.63.\tabularnewline
\bottomrule
\end{tabular*}
\end{small}
\end{table}

\subsection{Laser system}
The laser system \cite{LaserPaper} comprises four lasers: 369\,nm, 760\,nm, 798\,nm, and 935\,nm. Of these lasers, the 935\,nm and 760\,nm are commercially available as distributed feedback (DFB) devices as they are used in gas sensing applications. The 798\,nm is a volume holographic grating (VHG) extended cavity diode laser (ECDL) \cite{Rauch2015} and is frequency doubled to 399\,nm in a periodically-poled nonlinear waveguide.
The 369\,nm was provided by an ECDL.

The laser beams are transmitted from the laser system by polarization-maintaining (PM) fibers to fiber launchers that are fixed to the vacuum chamber. These
send the UV beam down the trap axis and the IR at 60\degree\,  to
the axis. Both beams are focused to a waist at a position after the ions.  At the ions, the $1/e$ intensity radius is $\unit[~200]{\mu m}$ for the 369-nm laser and $\unit[~300]{\mu m}$ for the 935-nm laser.

Although it is advantageous for the beams to travel down the trap axis as it ensures
the best overlap with the ions, it was simpler to send the IR and UV beams through separate windows.

\subsection{Electronics and microwave package}

The electronics system comprises the power supplies, microwave frequency
generation and experiment control. NPL-designed low noise current
sources are used for both the $\unit[369]{nm}$ laser and the Helmholtz coils.  These current sources are a modification of a Hall-Libbrecht design \cite{HallCS,Current_driver}. The current sources that are used for the other lasers are commercial devices. The lasers are
temperature controlled by an NPL-designed temperature controller.
The RF trap drive is based on a circuit developed by Sandia \cite{Jau2011}.

A field-programmable-gate-array (FPGA) controls the experimental sequence. A graphical user interface (GUI) allows the experimental sequences and timings to be altered from a laptop computer.

\begin{figure*}
\includegraphics[ width=1\textwidth]{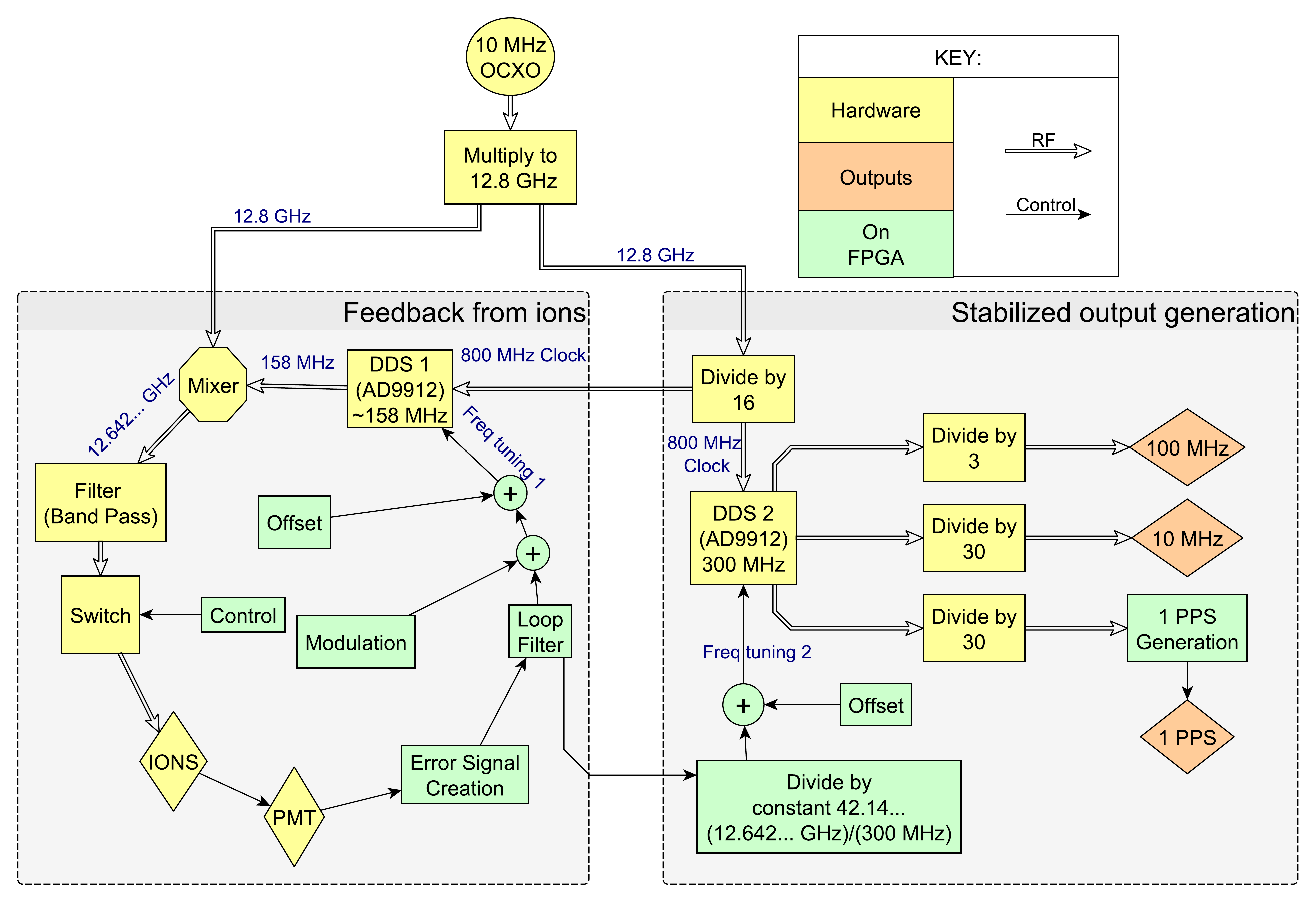}\caption{ Schematic of the frequency chain. The frequency chain produces both the 12.6\,GHz for driving the clock transition and the stabilized clock outputs from a free-running 10\,MHz LO.} \label{fig:FreqChain}
\end{figure*}

 The frequency chain to produce the 12.6\,GHz clock frequency from a 10\,MHz quartz (Rakon HSO 13) local oscillator is shown in Figure \ref{fig:FreqChain}. This frequency chain
multiplies a 10\,MHz input up to 12.8\,GHz. This 12.8\,GHz
signal is split into two paths, the first path is divided by 16 to clock two direct digital synthesizers (DDSs) (Analog Devices AD9912) and the second path is sent into a mixer. The second input of the mixer is a signal of around 158\,MHz that is used to generate the clock frequency as the low-frequency sideband of the mixer output. The 12.8\,GHz and the high-frequency sideband are then removed by a band-pass filter, and the clock frequency is then applied to the ions via an antenna.

The power of the applied microwaves can be controlled in the switching network of two single-pole-double-throw (SPDT) switches which provide up to 130\,dB of dynamic range between the high-power state, used for laser cooling, and the low-power state, used for state preparation and probing the clock-transition.

An error signal is derived through modulating the probe frequency to measure the high- and low- frequency sides of the resonance, and the frequency offset between
the free-running frequency chain and the ions is determined. This
frequency offset is used for both locking 12.6\,GHz to the ions and for providing the stabilised RF outputs.

The stabilised RF outputs are generated by updating DDS\,2 (Figure \ref{fig:FreqChain}) that outputs 300\,MHz. The frequency corrections
that are calculated by the feedback loop at 12.6\,GHz are divided
by a constant ($\unit[12.6418...]{GHz}/\unit[300]{MHz}\approx 42.14$) to ensure the appropriate fractional
change. This removes the drift from the free-running frequency chain
and transfers the stability from the ions' frequency
discriminant to the clock output. The AD9912 DDSs have a resolution
of $\unit[4]{\mu Hz}$, which means that the RF outputs can be tuned
with a fractional-frequency resolution of $1.4\times10^{-14}$, the
fractional frequency resolution of the probe frequency at 12.6\,GHz
is $4\times10^{-16}$.

This all-digital method relaxes the requirements for the tuning characteristics of the LO, and means designing a low-noise analogue voltage-tuning circuit is not required.

\section{Method of operation\label{sec:Meth}}
\sloppy
At the beginning of a measurement the lasers are frequency-stabilized by locking  to a   wavemeter that is external to the 6U 19-inch rack  \cite{Barwood2012} and a few hundred ions are loaded by passing a current through an oven in the presence of 399\,nm and 369\,nm light.
 The LO is then locked  to a Rabi resonance from the trapped ions.
The measurement cycle consists of four stages: laser cooling, state preparation,
clock pulse, and state-detection.

\begin{figure*}
\includegraphics[ width=1\textwidth]{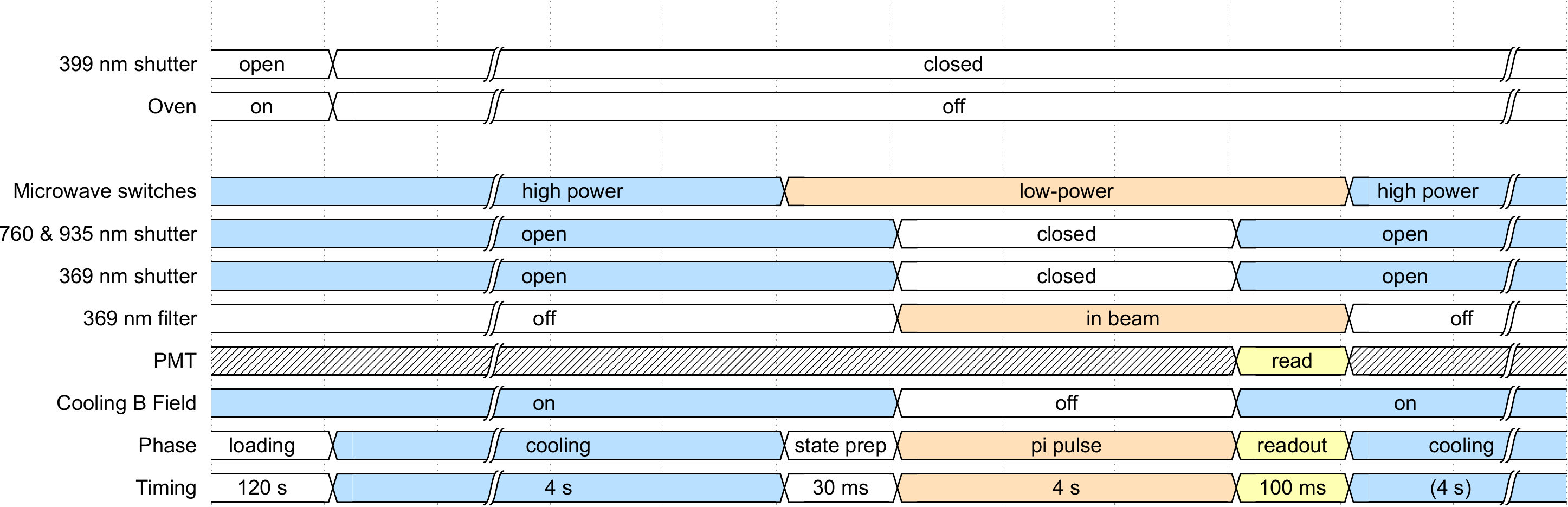}\caption{Clock measurement sequence, showing the states of the hardware elements during the measurement cycle, and the timings used for the measurements presented. The low-power microwaves are attenuated by 130\,dB from the high-power state.  The cycle time of $\approx\unit[8]{s}$ is dominated by the cooling and clock pulse times.} \label{fig:Sequence}
\end{figure*}

During the state-detection stage, a  neutral density filter is moved into the 369-nm beam to reduce the intensity to below the saturation level. This has the dual benefits of both reducing the number of  photons scattered off the compact trap and vacuum apparatus and also increasing the number of signal-photons by suppressing the off-resonantly driven $^{2}\text{S}_{\nicefrac{1}{2}}(\mathrm{F=1})\,\rightarrow\,6p:\,^{2}\text{P}_{\nicefrac{1}{2}}(\mathrm{F'=1})$ non-cycling transition. The contrast of the detected feature is also limited by incomplete state preparation that results from the $^{2}\text{S}_{\nicefrac{1}{2}}(\mathrm{F=0})\,\rightarrow\,6p:\,^{2}\text{P}_{\nicefrac{1}{2}}(\mathrm{F'=1})$ transition being driven at 10\% of the rate of the resonant  $^{2}\text{S}_{\nicefrac{1}{2}}(\mathrm{F=1})\,\rightarrow\,6p:\,^{2}\text{P}_{\nicefrac{1}{2}}(\mathrm{F'=0})$ transition. It is thought that this large off-resonant transition rate is due to amplified spontaneous emission from the 369-nm ECDL, this effect has been reported for similar 393-nm UV laser that is used in calcium ion systems \cite{Toyoda2001}. 
\color {black} The cycle time is dominated by the clock pulse and laser cooling, which for the data presented were both 4\,s in duration.
Figure \ref{fig:Sequence} shows the states of the apparatus during each of these stages of the measurement.

A linear drift-insensitive digital integrator feedback scheme is used
to find the new frequency of the offset of the LO, $f_{i}$, from the previous
frequency, $f_{i-1}$, and the fluorescence imbalance between the low and high-frequency 
probe frequencies $(S_{L}-S_{H})$. The two probe frequencies are
offset from $f_{i-1}$ by the same amount in the low- and high- frequency sides of the resonance. The gain, $g$, is calculated from a fit to the Rabi resonance. The feedback loop
is made insensitive to a linear signal-level drift by probing the
high and low-frequency sides of the resonance twice before applying
a correction, in a pattern: low, high, high, and then low:
\begin{equation}
f_{i}=f_{i-1}+\frac{g}{2}(S_{L1}+S_{L4}-S_{H2}-S_{H3}).\label{eq:freq_correction}
\end{equation}
The gain is divided by two in Equation \eqref{eq:freq_correction} because both
sides of the line are sampled twice before a correction is made.

\section{Results}\label{results}
\begin{figure}
\includegraphics[width=1\columnwidth]{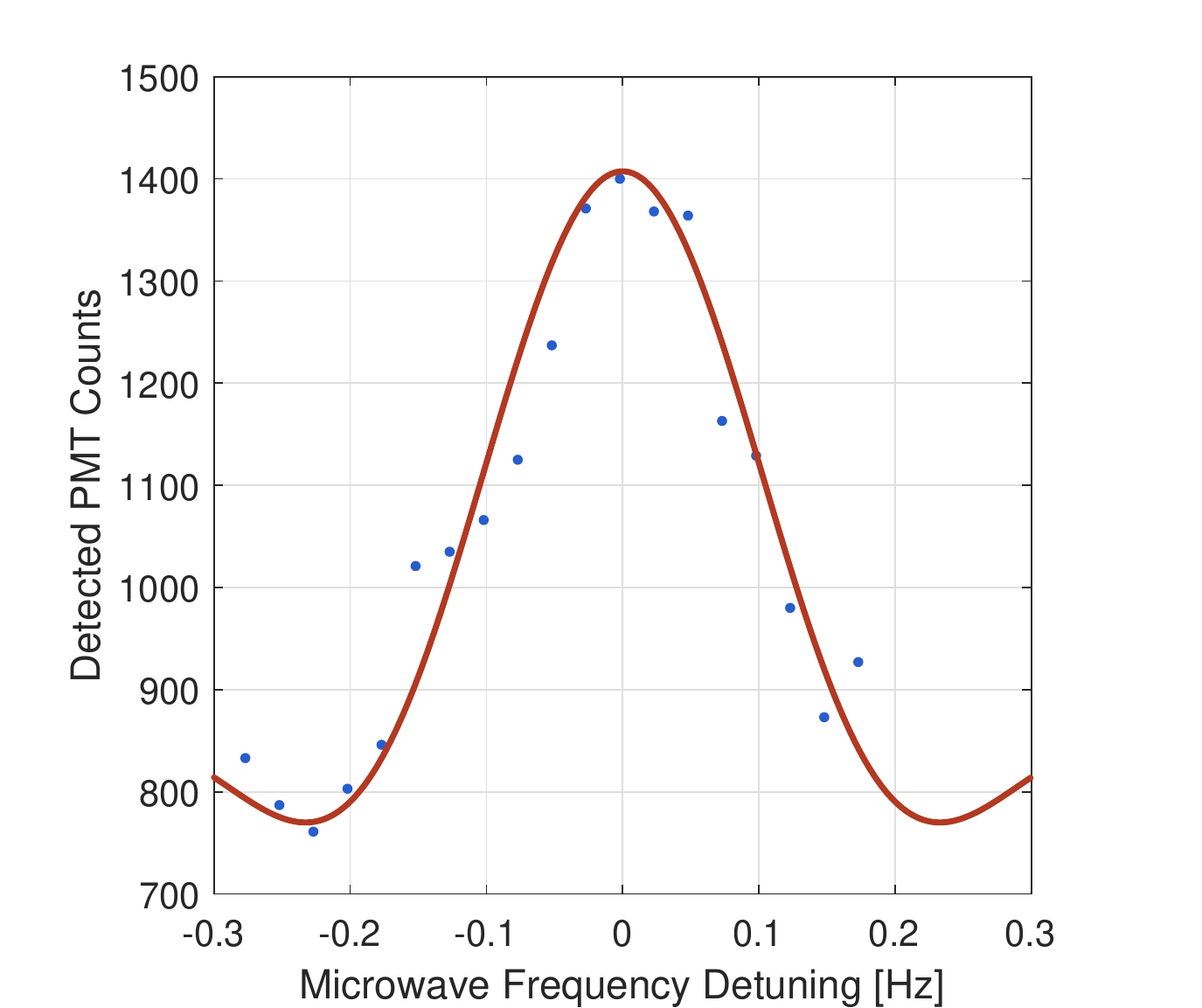}\caption[Microwave clock stability]{\label{fig:Rabi} Measured Rabi lineshape. The data were obtained from a single scan with a microwave pulse length of $\unit[4]{s}$ and detection time $\unit[100]{ms}$. The frequency scale is relative to the peak of the resonance, offset from the unperturbed frequency by about 2\,Hz. These data were taken at the beginning of the stability measurement.}
\end{figure}

A Rabi lineshape was measured by sweeping the microwave frequency through the resonance, and this is shown in Figure \ref{fig:Rabi}. Following the frequency sweep, the microwaves were then locked to the resonance. The lock to the clock transition produces a stabilized 10\,MHz output signal and this was compared with the 10\,MHz signal from a maser via a phase comparator (Microsemi 3120A). The phase difference was logged as a function of time, and the Allan deviation was calculated and displayed in real time on a PC.  

\begin{figure}
\includegraphics[width=1\columnwidth]{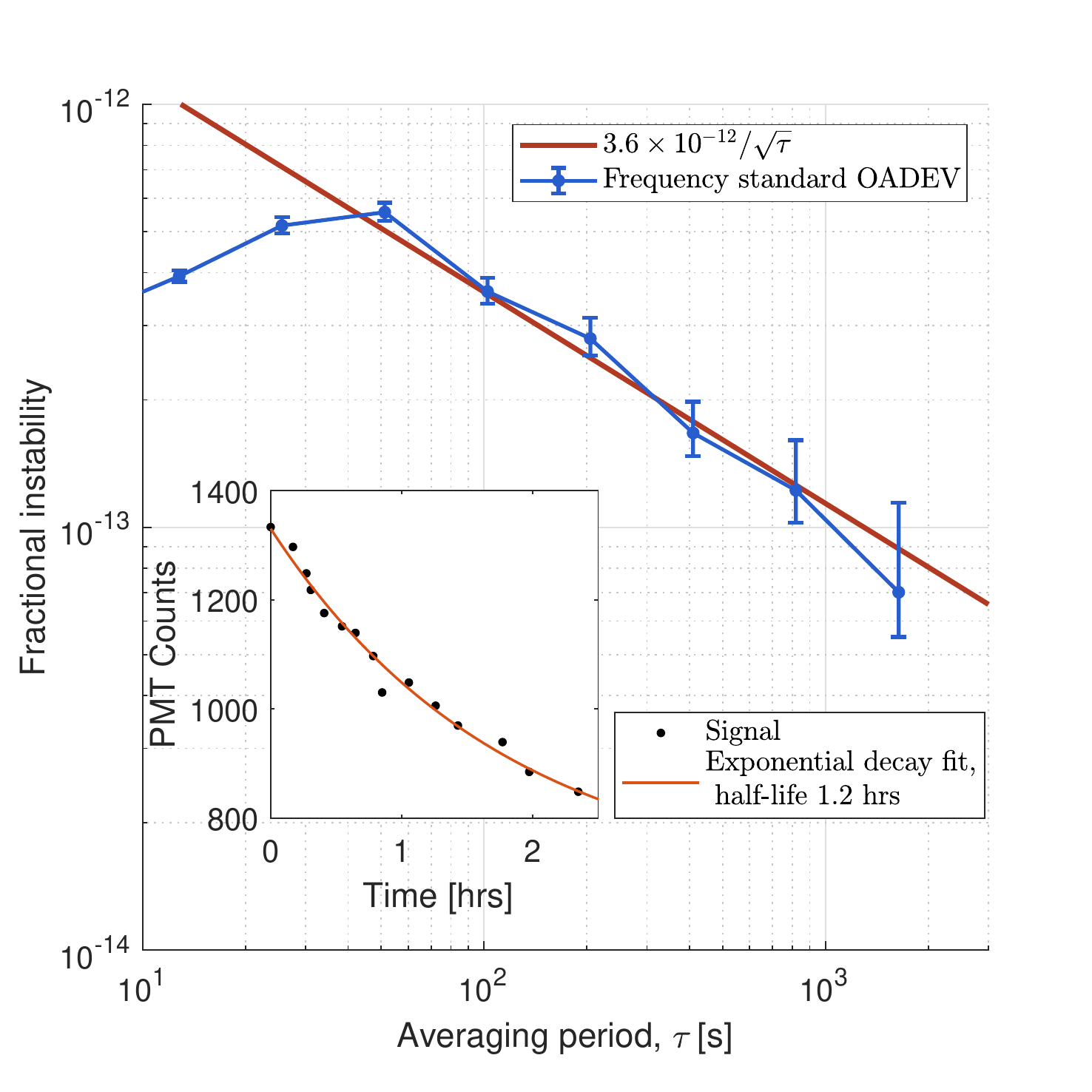}\caption[Microwave clock stability]{\label{fig:Stability-3.7e-12}Stability of the frequency standard measured
against a hydrogen maser. The fractional overlapping Allan deviation (OADEV) is plotted against averaging time. An instability of $3.6\pm0.2\times10^{-12}/\sqrt{\tau}$
is measured between averaging times of 30\,s to 1500\,s. Signal loss, shown in the inset, limits the data to  $2\nicefrac{1}{4}$\,hrs.}
\end{figure}

Figure \ref{fig:Stability-3.7e-12} shows frequency stability data taken over a continuous period of $2\nicefrac{1}{4}$\,hrs. During the measurement, the signal level was decreasing with a half-life of 1.2\,hrs. This resulted in increased instability from shot-noise as discussed in the following section. An instability of $3.6\pm0.2\times10^{-12}/\surd\tau$  
for averaging times between $\unit[30]{s}$ and $\unit[1500]{s}$ was measured.  

The following sections detail the causes of instability that affect the performance of this frequency standard. These include shot-noise when measuring the clock-transition, which limits the stability for the data presented, and instabilities due to changing environmental parameters that must be controlled for long-term stability performance.
The results of this analysis are summarized in Table \ref{tab:table3}.

\subsection{Shot noise and dead time\label{subsec:SN}}
 From the scan over the Rabi feature that was taken at the beginning of the measurement, the shot-noise limit to the instability was estimated as  $\sigma_{\mathrm{SN}}(\tau)\approx1\times10^{-12}/\sqrt{\tau}$
 by using the formula \cite{SNRQ}: \begin{equation}
 \sigma_{\mathrm{SN}}(\tau)\approx\frac{\sqrt{\tau_p + \tau_d}}{\pi \cdot \mathrm{SNR}\cdot Q}\times\frac{1}{\sqrt{\tau}}.\end{equation}The clock-transition was probed for $\unit[\tau_p=4]{s}$ and there was $\unit[\tau_d=4]{s}$ of cooling time between measurements. The signal-to-noise ratio at the peak ($\mathrm{SNR}$) was estimated as $16$ by using the formula:
 \begin{equation}
 \mathrm{SNR}\approx\frac{\mathrm{SH}}{\sqrt{\mathrm{BG}+\mathrm{SH}}},\end{equation}where $\mathrm{SH}$ is the signal level above the background at the maximum point and $\mathrm{BG}$ is the background signal level  that arises from light scattered from the ion trap electrodes, and incomplete state preparation. 
 
 The quality factor ($\mathrm{Q}$) was $6.32\times10^{10}$ from the FWHM of 0.2\,Hz and the transition frequency of 12.6\,GHz. 
 Signal loss caused the $\mathrm{SNR}$ to decrease exponentially  over time with a half-life of $\unit[1.2]{hrs}$ and so the instability was higher than that estimated from the initial Rabi resonance. It is believed that this signal loss is due to molecular ions forming by the ytterbium ions chemically reacting with gases that are released into the vacuum system when the ytterbium source is heated.

 The pressure in the vacuum system was measured by monitoring the current drawn from the ion pump controller.  These currents are a few nanoamperes and there is a large uncertainty in the derived pressure. During clock operation, the reported pressure was below $\unit[10^{-10}]{mbar}$ but when the oven is heated to load ions into the trap, this rises to  $\unit[1\times10^{-9}]{mbar}$.
 
  At $\unit[<10^{-10}]{mbar}$, the expected ion-neutral collision rate is 15 per ion per hour using the Langevin model for ion-neutral atom collisions  \cite{CharlesThesis,Tomza2018}. Ytterbium single-ion optical clocks have shown ion lifetimes greater than one month \cite{King2012} but ytterbium ions are known to react rapidly with $\mathrm{H_{2}O}$ and $\mathrm{O_{2}}$ molecules \cite{Sugiyama1995} and these gases may be emitted when the oven is heated. A new oven design with a lower operating current and improved thermal isolation is undergoing tests and this shows much lower levels of fluorescence loss. 

\subsection{Second-order Zeeman shift}\label{subsec:SOZ}

As the clock transition has no linear Zeeman shift at zero magnetic field, the frequency sensitivity to magnetic fields is primarily from the second-order Zeeman effect. For $\mathrm{^{171}Yb^{+}}$ this is \cite{Vanier:1989}:
\begin{equation}
\frac{\Delta\nu_{2OZ}}{B^2}=31.08\,[\mathrm{mHz/(\mu T)^{2}}].
\end{equation}
Small DC fluctuations, $\mathbf{\delta B}$, on a background magnetic field, $\mathbf{B}$, result in changes to the second-order Zeeman shift, $\delta\nu_{2OZ}$, given by:
\begin{equation}\label{eqn:SOZ}
\frac{\delta\nu_{2OZ}}{\nu_{clock}}=4.92\times10^{-12}\,[\mu T^{-2}]\,\mathbf{B}\cdot\mathbf{\delta B} ,
\end{equation}
to first order in $\mathbf{\delta B}$.

A stable magnetic field giving rise to a constant second-order Zeeman shift is not a problem, provided that it is corrected for; fluctuating magnetic fields provide a greater challenge. To mitigate this effect, the scalar product of the terms $\mathbf{B}$ and $\mathbf{\delta B}$ is minimized by making these vectors orthogonal.
$\mathbf{B}$ is reduced by adjusting the currents in three orthogonal sets of Helmholtz coils to null the background magnetic field. Subsequently, a small and well-controlled bias of $\unit[8]{\mu T}$ is applied to define the quantization-axis. At this bias field the second-order Zeeman shift is about $\unit[2]{Hz}$, and to first order, the fractional frequency sensitivity to magnetic field fluctuations parallel to the bias field is $\unit[4.0\pm0.1\times10^{-11}]{/\mathrm{\mu T}}$.
The $\mathbf{\delta B}$ term is minimized by operating the trap in a magnetic shield with a  DC\, shielding factor parallel to the bias field of $40\pm4$  (and approximately $750$ and $20$ in the two perpendicular directions), and by using low noise power supplies to drive the Helmholtz coils.

Both the background magnetic field in the laboratory, and the current passing through the quantization-axis coils, were measured to determine the instability of the second-order Zeeman shift. The magnetic field in the laboratory was measured with a fluxgate magnetometer, outside the magnetic shields, a few centimeters away from the ion trap. The coil current was measured across a $\unit[20]{\Omega}$ precision resistor on a  voltmeter. 

\begin{figure}
\centering{}\includegraphics[width=1\columnwidth]{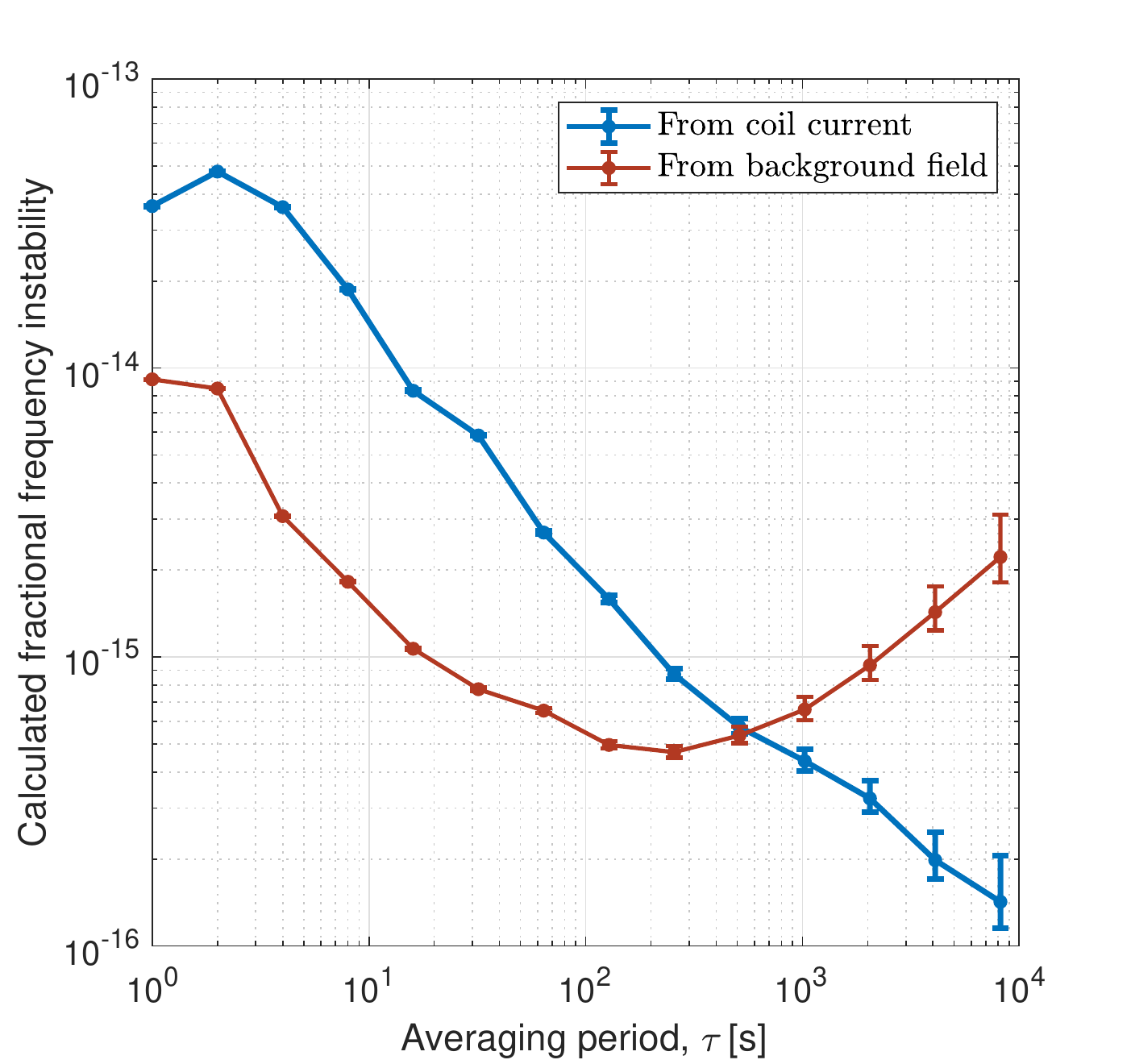}\caption[Temperature measurement of the ion cloud after a 4\,s clock pulse]{\label{fig:SOZ} Calculated second-order Zeeman instability from the measured fluctuations in both the background magnetic field of the laboratory (red trace) and the magnetic field from the bias coils (blue trace). The drift in the background field that is evident after a few hundred seconds could be removed or corrected for by a control system. }
\end{figure}

The instability in the second-order Zeeman shift is shown on Figure \ref{fig:SOZ} and the results are summarized in Table \ref{tab:table3}. The instability due to fluctuating magnetic fields is small compared to the instability from the shot-noise. There is also an AC second-order Zeeman shift that arises from oscillating magnetic fields \cite{ACZeemanPRA}. The oscillating magnetic fields are primarily from two sources, $\unit[50]{Hz}$ mains AC and the RF trap drive. The $\unit[50]{Hz}$ field was measured using the fluxgate magnetometer. This sensor \color{black} is sensitive to oscillations below a few hundred Hertz and an oscillating magnetic field with amplitude $\unit[10]{\mu T}$ at $\unit[50]{Hz}$ was measured parallel to the quantization axis. Assuming that the $\unit[50]{Hz}$ shielding factor is similar to the DC value, this results in a frequency shift of $8\pm1 \times10^{-14}$ and an instability that is negligible over the timescales of our frequency stability measurement.

The trap capacitance causes small oscillating currents at the trapping frequency in the trap rods. For a perfectly symmetric trap, the magnetic fields generated by these currents will cancel along the nodal line of the trapping field. However, if the ions are not centered, or if  the currents are not symmetric, an oscillating magnetic field will cause an AC Zeeman shift \cite{ACZeemanPRA,Berkeland1998merc}.
The fractional frequency shift from this effect was calculated to be below $1\times10^{-15}$.

\color{black}

\subsection{Second-order Doppler shift\label{subsec:SOD}}
The frequency shift due to the second-order Doppler effect is given
by: 

\begin{equation}
\Delta\nu_{2OD}=-\frac{\nu}{2}\times\frac{u^{2}}{c^{2}},
\end{equation}
where $\nu$ is the center frequency of the resonance, $u$ is the velocity of the ion and $c$ is the speed of light.

\begin{figure}[t]
\centering{}\includegraphics[width=1\columnwidth]{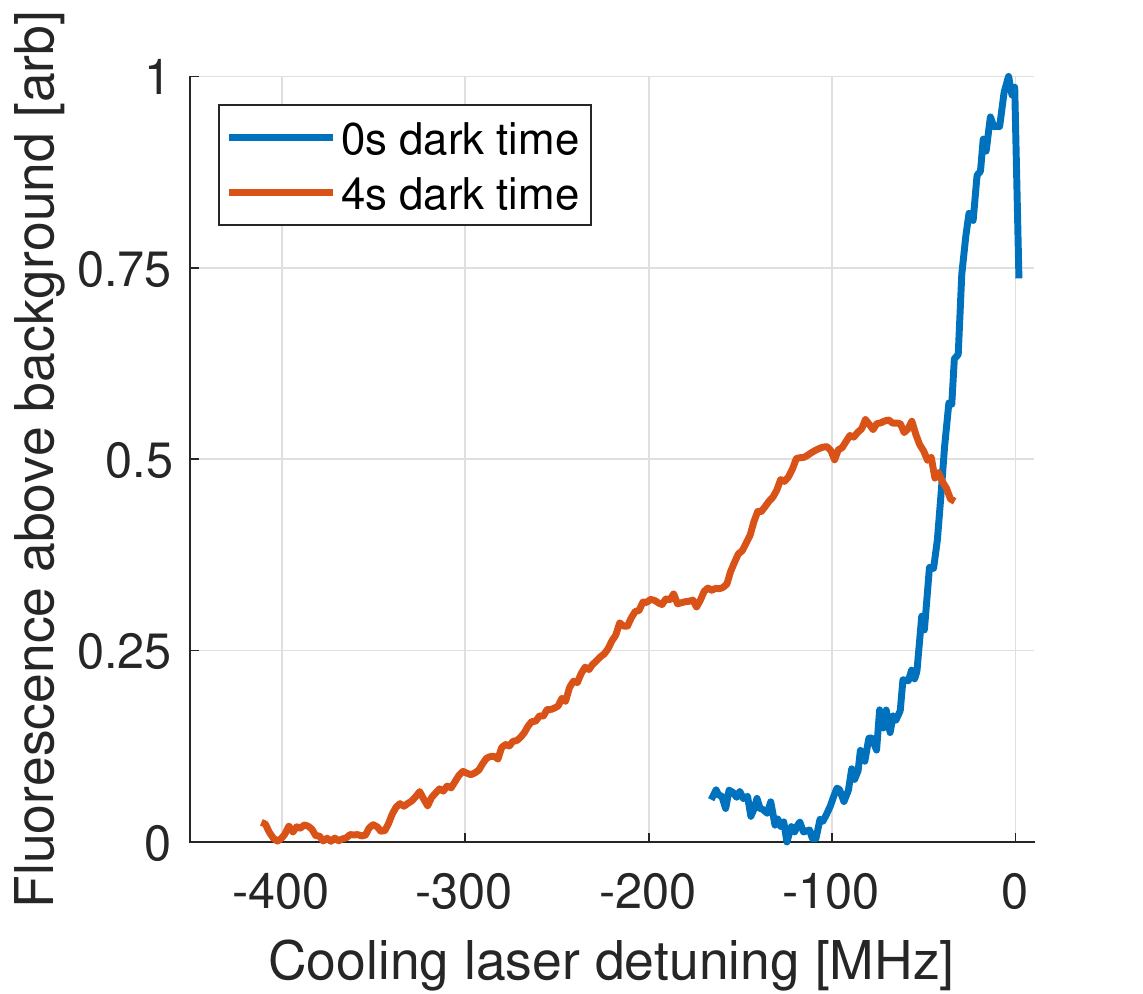}\caption[Temperature measurement of the ion cloud after a 4\,s clock pulse]{\label{fig:SOD}Doppler broadened profiles of the laser-cooling transition.
The trace in blue shows the fluorescence profile with no dark time
and can be used to infer a rms velocity of $\unit[18\pm2]{m/s}$. Each of the points
on the orange trace is a measurement immediately after a 4\,s period
where the cooling laser was blocked. To avoid cooling the ions, the
measurements were taken within 10\,ms of the cooling laser becoming
unblocked. From the profile of the transition, the rms velocity after the clock pulse is calculated to be $\unit[55\pm10]{m/s}$. These rms velocities are likely to be
over-estimates for reasons that are explained in the text. The y-axis scale is relative to the peak fluorescence with no dark time, the background from light that is not scattered by the ions is removed.}
\end{figure}

A measurement of the ion velocities was made by observing the first-order Doppler shift on the $\unit[369]{nm}$ cooling transition. The laser detuning was scanned from the red side of the resonance towards line center both when the ions were continuously cooled and when the laser was extinguished for $\unit[4]{s}$ between measurements to simulate a clock measurement. The results are shown in Figure \ref{fig:SOD}. As may be expected from a cloud of ions, it is evident that there is significant micromotion \cite{DeVoe1989,Berkeland_micromotion}. To calculate the second-order Doppler shift, the root mean square (rms) velocity was calculated by numerical integration of the velocity distributions implied from these scans over the cooling transition. These calculated velocities are over-estimates as the method assumes that all the broadening of the transition is due to first-order Doppler shifts and because laser heating effects have suppressed the fluorescence of the ions at small detunings. Between measurements the laser is not returned to line-center and the effect of this incomplete recooling will also contribute to an over-estimate of the rms velocity. From these measurements the second-order Doppler shift was estimated to be below $-2.0\pm0.5\times10^{-14}$ and have an instability less than $5\times10^{-15}$.

\subsection{Quadratic Stark shift\label{subsec:QSS}}

The quadratic Stark shift from the RF trapping field is given by \cite{Fisk1997}:
\begin{align}
\frac{\Delta\nu_{AC\,Stark}}{\nu_{clock}} & =-k_{s}\left\langle \boldsymbol{E}^{2}\right\rangle \\
 & =-4k_{s}\left(\frac{\omega_{single}}{\omega_{cloud}}\right)^{2}\frac{m\Omega_{trap}^{2}}{e^{2}}k_{B}T,
\end{align}
%\sloppy
where $e$ is the elementary charge and $\Omega_{trap}$ is the RF
trap frequency. The value of the coefficient $k_{s}$ has been measured
experimentally \cite{Tamm1995}  to be  $k_{s}=2 \pm$ \linebreak $1 \times10^{-21}\,\mathrm{m^{2}V^{-2}}$.
 The trap drive frequency was $\unit[(2\pi)\times2]{MHz}$,
and the peak-to-peak voltage was $\unit[300]{V}$. The ratio of the single-ion and cloud secular frequencies $\frac{\omega_{single}}{\omega_{cloud}}$ was measured to be $1.0\pm0.1$. The resulting fractional frequency offset is $2\pm1\times10^{-15}$ and instability from the quadratic Stark shift is at the $10^{-15}$ level.
%\fussy
\subsection{Black-body radiation (BBR) shift (AC Stark shift)\label{subsec:BBR}}

There is an AC Stark shift due to the thermal radiation produced by
the ions' surroundings, $T$, given by \cite{Angstmann2006}:

\begin{equation}
\frac{\Delta\nu_{BB}}{\nu_{clock}}=\beta\left(\frac{T}{T_{0}}\right)^{4}\left[1+\epsilon\left(\frac{T}{T_{0}}\right)\right]^{2},\label{eq:blackbody}
\end{equation}where $T_{0}$ is the room temperature and 300\,K and the coefficients are \cite{Angstmann2006,Safronova2009} $\beta=-9.83\times10^{-16}$, and $\epsilon=2\times10^{-3}$.

An estimate of the fractional frequency shift from BBR
of $-1.0\pm 0.1\times10^{-15}$ is obtained by assuming a 20\,K
range in temperature for the ions' surroundings, from 300\,K
to 320\,K. A fractional frequency sensitivity to the temperature
close to 310\,K is calculated to be $\unit[-1\times10^{-17}]{/K_{T\approx310\thinspace K}}$.
Assuming an unstable temperature at the 10\,K level gives an
estimate of the fractional frequency instability of the BBR shift
of $<1\times10^{-16}$.

\subsection{Dick effect\label{subsec:DE}}

In the measurements reported in this paper, the duty cycle ---
defined as the fraction of the measurement devoted to probing the transition
--- was $\nicefrac{1}{2}$. Due to the ions not being continuously
measured, some high-frequency noise on the oscillator will be aliased
to lower frequencies and will add instability to each clock measurement.
The instability added for a duty cycle of $\nicefrac{1}{2}$ was calculated
by Dick \cite{Dick1987} as 1.4 times below the local oscillator instability
at the measurement time. As the instability from shot-noise from the ion signal was greater than the instability from the LO at the measurement time, we are not limited by the Dick effect.

\begin{table*}
\caption[ Sources of instability and frequency offsets]{\label{tab:table3} Summary of the sources of noise, frequency uncertainty and frequency offsets during this frequency stability measurement. Further details are within the referenced sections of the text.}
\noindent
\begin{center}
{\footnotesize{}}%
\begin{tabular}{>{\raggedright}m{0.15\textwidth}>{\raggedright}m{0.3\textwidth}>{\raggedright}m{0.17\textwidth}>{\raggedright}m{0.15\textwidth}>{\raggedright}m{0.17\textwidth}}
\multirow{2}{\textwidth}{\centering{}\textbf{\scriptsize{}}} & \multirow{2}{\textwidth}{\centering{}\textbf{\scriptsize{}}} & \multicolumn{3}{c}{\textbf{\scriptsize{}Fractional frequency:}}\tabularnewline 
\centering{}\textbf{\scriptsize{}Effect} & \centering{}\textbf{\scriptsize{}Parameter values} & \centering{}\textbf{\scriptsize{}Sensitivity} & \centering{}\textbf{\scriptsize{}Offset} & \centering{}\textbf{\scriptsize{}Instability}\tabularnewline\hline
{\scriptsize{}Shot-noise and dead time} & {\scriptsize{} 650 initial signal and 750
background photons, duty cycle $\nicefrac{1}{2}$,
signal~loss~with~half-life~1.2\,hrs} 
{\scriptsize\par}
{\scriptsize{} [sec: \ref{subsec:SN}] }
& \centering{}--- & \centering{}--- & {\scriptsize{}$3.6\pm0.2\times10^{-12}/\sqrt{\tau}$ }{\scriptsize\par}
{\scriptsize{}$30<\tau<1500$ s}
\tabularnewline
\hline 
{\scriptsize{}Second-order Zeeman: coils} & {\scriptsize{}$\unit[B_{bias}=8.0\pm0.2]{\mu T}$}
{\scriptsize\par}
{\scriptsize{} [sec: \ref{subsec:SOZ}] } & {\scriptsize{}$\unit[4.0\pm0.1\times10^{-11}]{/\mathrm{\mu T}}$ } & {\scriptsize{}$1.60\pm0.08\times10^{-10}$\normalsize{*}} & {\scriptsize{}$<1\times10^{-13}$}{\scriptsize\par}

{\scriptsize{}$1<\tau<10^{2}$ s,}{\scriptsize\par}

{\scriptsize{}$1.4\pm0.1\times10^{-14}/\sqrt{\tau}$}{\scriptsize\par}

{\scriptsize{}$10^{2}<\tau<10^{4}$ s}%\tabularnewline
{\scriptsize\par}
{\scriptsize{}[fig: \ref{fig:SOZ}]}\tabularnewline
\hline 
{\scriptsize{}Second-order Zeeman: background DC } & {\scriptsize{}$B_{lab}\approx\unit[50]{\mu T}$\par Shielding factor $=40\pm4$}
{\scriptsize\par}
{\scriptsize{} [sec: \ref{subsec:SOZ}] } & {\scriptsize{}$\unit[1.0\pm0.1\times10^{-12}]{/\mathrm{\mu T}}$} & {\scriptsize{}nulled by coils} & {\scriptsize{}$<1\times10^{-14}$}{\scriptsize\par}

{\scriptsize{}$1<\tau<10^{4}$ s}%\tabularnewline
{\scriptsize\par}
{\scriptsize{}[fig: \ref{fig:SOZ}]}
\tabularnewline
\hline 
{\scriptsize{} AC Zeeman } & {\scriptsize{}$B_{\unit[50]{Hz}}<\unit[0.25]{\mu T}$\par $B_{\Omega}\unit[<20]{nT}$ }
{\scriptsize\par}
{\scriptsize{} [sec: \ref{subsec:SOZ}] } & {\scriptsize{}$\unit[2.5\times10^{-12}]{/(\mathrm{\mu T}})^2$} & {\scriptsize{}$8\pm1\times10^{-14}$ ($\unit[50]{Hz}$) \par $<1\times10^{-15}$ ($\Omega$)} & {\scriptsize{}$<1\times10^{-15}$}
\tabularnewline
\hline 
{\scriptsize{}Second-order Doppler} & 
{\scriptsize{}$v_{\mathrm{rms}} \approx\unit[18]{m/s}\rightarrow\unit[55]{m/s}$ }
{\scriptsize\par}
{\scriptsize{} [sec: \ref{subsec:SOD}] }& 

{\scriptsize{}$\unit[-5.6\times10^{-18}]{/v_{rms}^2}$} & {\scriptsize{}$-2.0\pm0.5\times10^{-14}$} & {\scriptsize{}$<5\times10^{-15}$}
\tabularnewline
\hline 
{\scriptsize{}Quadratic Stark shift: trap drive} & {\scriptsize{}$\Omega=\unit[2]{MHz}$, $V_{pk-pk}=\unit[300]{V}$}
{\scriptsize\par}
{\scriptsize{} [sec: \ref{subsec:QSS}] }& {\scriptsize{}$\unit[<1\times10^{-16}]/{V_{pp}}$} & {\scriptsize{}$2\pm1\times10^{-15}$} & {\scriptsize{}$<1\times10^{-15}$}

\tabularnewline
\hline 
{\scriptsize{}BBR shift} & {\scriptsize{}$\unit[300\leq T_{\mathrm{trap}}\leq320]{K}$}
{\scriptsize\par}
{\scriptsize{} [sec: \ref{subsec:BBR}] }& {\scriptsize{}$\unit[-1\times10^{-17}]{/K_{T\approx310\thinspace K}}$} & {\scriptsize{}$-1.0\pm0.1\times10^{-15}$} & {\scriptsize{}$<1\times10^{-16}$}\tabularnewline
\hline 
{\scriptsize{}Dick effect} & {\scriptsize{}duty cycle $\nicefrac{1}{2}$, Feedback
at 30\,s} {\scriptsize\par}
{\scriptsize{} [sec: \ref{subsec:DE}] }
& \centering{}--- & \centering{}--- & {\scriptsize{}$\sim1\times10^{-13}/\sqrt{\tau}$}\tabularnewline
\hline 

\end{tabular}{\footnotesize\par}

\par\end{center}
% hacked table footnote
* \hangindent=0.3cm This uncertainty is due to the method used to measure the second-order Zeeman shift during this stability measurement. For a full-accuracy implementation of our clock, we would monitor this shift using the $\mathrm{F=1,m_{F}=\pm1}$ magnetically sensitive Zeeman components of the clock transition and this uncertainty could be reduced to below $1\times10^{-14}$.

\end{table*}

\section{Future Developments    \label{sec:Disc}}

\begin{figure}[]

\includegraphics[ width=1\columnwidth]{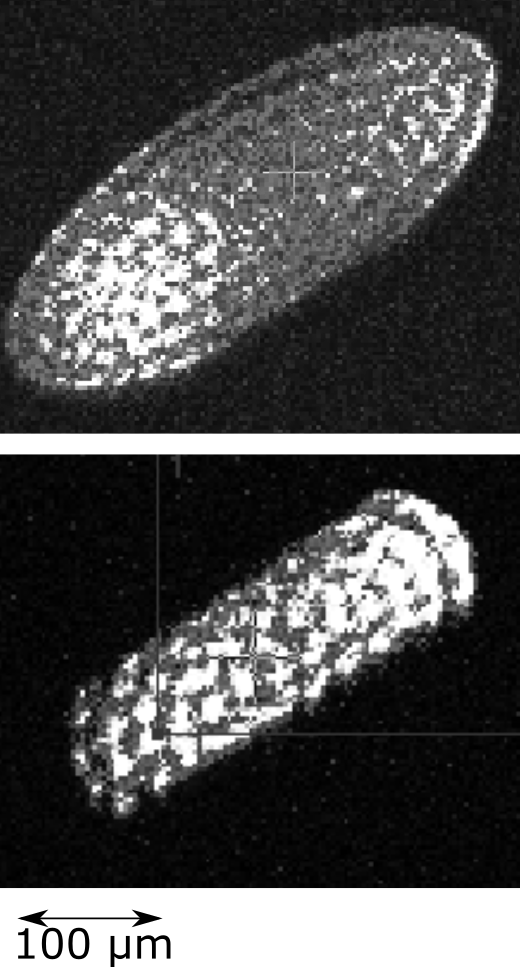}\caption{ CCD images of ion crystals. The top image shows a crystal that contains only $^{171}\mathrm{Yb}^{+}$ ions and was taken shortly after loading. The lower image shows the same sample but a few hours later. It can be seen that the remaining $^{171}\mathrm{Yb}^{+}$ ions are confined to the center of the original crystal outline. The distorted shape of the fluorescing ions is due to heavier, non-fluorescing ions, occupying the lattice sites that complete the prolate spheroid shape of the minimum-potential volume.   \label{fig:dark-ion}}

\end{figure}

The most significant limitation to the performance of the prototype described in this paper arises from signal loss via an accumulation of non-fluorescing (dark) ions that  appear to be molecular ions formed by $^{171}\mathrm{Yb}^{+}$ reacting with background gas. Attempts to recover fluorescence using either $\unit[760]{nm}$ light at F state repump frequencies that we measured from a $\mathrm{Yb}^{+}$ optical clock  \cite{LaserPaper}  ($\unit[760.065(1)]{nm}$ and $\unit[760.075(1)]{nm}$), or the $\mathrm{YbH}^{+}$ dissociation wavelength --- 369.482 nm \cite{Sugiyama1997},  were not successful. The observation that the dark ions accumulate primarily on the radial edges of a Coulomb crystal as seen in Figure \ref{fig:dark-ion} implies that the dark ions are heavier than $^{171}\mathrm{Yb}^{+}$ and so we conclude that they are most likely to be Yb molecular ions. This separation occurs because the central ion-trapping force is stronger for lighter ions than heavier ions \cite{Dubost2014, Hornekaer2001}.  Identification via molecular mass determination from the secular frequencies \cite{Sugiyama1995,Sheridan2011} is not straightforward on account of the large $^{171}\mathrm{Yb}^{+}$ mass.

 Investigations are ongoing as ion loss clearly impacts on both long-term frequency stability and operational time, and reduces the efficiency of the laser cooling. Before the dark ions form, it takes 500\,ms for the fluorescence to recover after the cooling laser is blocked for 4\,s during a clock pulse. This recovery time increases as the number of dark ions increases. To ensure consistent cooling we use a conservative cooling time of 4\,s.

The mounting and heatsinking of our oven are also undergoing design changes and preliminary off-line tests indicate significant performance improvements. Our new oven arrangement requires considerably less current and we believe that this will lead to reduced ion loss and more rapid ion re-loading in a future trap system. Reduced oven currents can also be expected to lead to reduced changes in ambient magnetic field on loading and so improve load-to-load frequency reproducibility. This short-term reproducibility (system re-trace) is currently limited at the $2\times10^{-13}$ level which we believe is dominated by temporary magnetization of between 10\,nT and 25\,nT generated by the oven current.

The dominant frequency shift for this clock arises from the second-order Zeeman effect and this is determined by our applied bias field $\mathbf{B} \approx \unit[8]{\mu T}$.
This high bias field is required because, in the presence of a magnetic field gradient, the clock transition can be broadened when the Zeeman splitting equals that of a secular frequency.
This is discussed in Ref. \cite{Partner2013} and creates an additional clear-out path from the  $ \mathrm{F} = 1, \Delta \mathrm{m_{F}} = 0$ clock level, broadening this transition.
Operating at a magnetic field giving a Zeeman splitting of more than the secular frequencies restricts us to $\mathbf{B} \geq \unit[8]{\mu T}$ and therefore sets a limit on the Zeeman shift (equation \eqref{eqn:SOZ}) and its contribution to the frequency stability arising from changes in the ambient magnetic field.
However, by incorporating a second lock to the $ \mathrm{F} = 1,\mathrm{m_{F}} = \pm 1$ magnetically sensitive Zeeman components, corrections for magnetic field variations could be applied, and this lock is intended for future implementation.

This section has outlined our plans for trap system improvements to improve device robustness, frequency stability and reproducibility. In particular, off-line tests of a new oven design indicate that significant improvements in performance are possible. Additionally, we discussed in Ref. \cite{LaserPaper} plans for the implementation of a more compact wavemeter and a possible alternative scheme to generate the $\unit[369.5]{nm}$  cooling radiation.

\section{Conclusions\label{sec:Conc}}

This paper describes the performance of a compact microwave frequency standard demonstrator based on laser-cooled ytterbium ions in a linear Paul trap. The short-term frequency instability achieved over averaging periods of 30\,s to 1500\,s  was $3.6 \times 10^{-12}/\sqrt{\tau}$ which is an improvement over that of the best commercial compact atomic clocks. 
A shot-noise limited instability of $1\times10^{-12}/\sqrt{\tau}$ is implied from our measurements of the initial lineshape. Enhancements to this system will address the fluorescence loss issue and allow longer-term operation.  The next design iteration will enable the reproducibility to be investigated at the $10^{-14}$ level and determine the limits to the ultimate accuracy achievable. Together with a substantial reduction in size, and further development of the laser system \cite{LaserPaper}, these improvements should allow demonstration of the full capability of our laser-cooled frequency standard.

\section*{Acknowledgements}
The Authors would like to gratefully acknowledge U.K. Defence Science and Technology Laboratory (Dstl) and Innovate UK for funding the development here reported. We would also like to thank our colleagues Pravin Patel for his assistance with the electronics, Peter Nisbet-Jones for his work on the vacuum system, and Steven King for his work at the beginning of the project.
%\end{acknowledgements}

%\bibliographystyle{spphys}
%\bibliographystyle{ieeetr}
%\bibliography{manual_lib_nourl.bib}

%\iffalse

%\fi

\end{document}